\documentclass[aps,prd,twocolumn,a4paper,final,superscriptaddress,longbibliography]{revtex4-1}
\usepackage{graphicx}
\usepackage{comment}
\usepackage{bm}
\usepackage{amsmath,amssymb}
\usepackage{ascmac}
\usepackage{ulem}

\newcommand{\Tr}{{\mathrm{Tr}}} 
\newcommand{\Det}{{\mathrm{Det}}}
\newcommand{\dd}{{\mathrm{d}}} 
\newcommand{\pd}{{\partial}} 
\newcommand{\DOne}{{\Delta_1}} 
\newcommand{\DTwo}{{\Delta_2}} 
\newcommand{\rmA}{\mathrm{A}} 
\newcommand{\rmB}{\mathrm{B}} 
\newcommand{\rmC}{\mathrm{C}} 
\newcommand{\ZA}{Z_\mathrm{A}} 
\newcommand{\ZB}{Z_\mathrm{B}} 
\newcommand{\ZC}{Z_\mathrm{C}} 
\newcommand{\n}{\mathfrak{n}}
\newcommand{\m}{\mathfrak{m}}
\newcommand{\vecn}{\vec{n}}
\newcommand{\Lag}{\mathcal{L}}
\newcommand{\integer}{\mathbb{Z}}
\newcommand{\Real}{\mathbb{R}}
\newcommand{\Complex}{\mathbb{C}}
\newcommand{\B}{\mathcal{B}}
\newcommand{\K}{\mathcal{J}}
\newcommand{\A}{\mathcal{A}}
\newcommand{\Z}{\mathcal{Z}}
\newcommand{\vecZ}{\vec{\mathcal{Z}}}
\newcommand{\bdb}{\beta^\dagger\dot{\beta}}
\newcommand{\veco}{\bm{\omega}}

\newcommand{\tri}{\text{tri}}
\newcommand{\nontri}{\text{nontri}}
\newcommand{\cl}{\text{cl}}
\newcommand{\nls}{nonlinear $\sigma$-model }

\def\({\left(}
\def\){\right)}
\def\[{\left[}
\def\]{\right]}
\def\<{\left\langle}
\def\>{\right\rangle}

\newcommand{\Kahler}{K\"{a}hler }
\newcommand{\Backlund}{B\"{a}cklund }
\newcommand{\ansatze}{ans\"{a}tze }

\begin{document}
	
	\title{$SU(3)$ Knot Solitons: Hopfions in the $F_2$ Skyrme-Faddeev-Niemi model }
	
	\author{Yuki Amari}
	\email{amari.yuki.ph@gmail.com}
	\address{Department of Physics, Tokyo University of Science, Noda, Chiba 278-8510, Japan}

	\author{Nobuyuki Sawado}
	\email{sawadoph@rs.tus.ac.jp}
	\address{Department of Physics, Tokyo University of Science, Noda, Chiba 278-8510, Japan}

	\vspace{.5 in}
	\small

	\date{\today}
	
	\begin{abstract}
	We discuss the existence of knot solitons (Hopfions)  in a  
	Skryme-Faddeev-Niemi-type model on the target space $SU(3)/U(1)^2$,
	which can be viewed as an effective theory of both the $SU(3)$ Yang-Mills theory  and
	the $SU(3)$ anti-ferromagnetic Heisenberg model.
	We derive the knot solitons with two different 
	types of ansatz: the first is a trivial embedding configuration of $SU(2)$ into $SU(3)$,  and the second 
	is a non-embedding configuration  that can be generated  through the \Backlund transformation. 
	The resulting Euler-Lagrange equations for both ansatz 
	reduce exactly to those of the $CP^1$ Skyrme-Faddeev-Niemi model. We also examine some quantum aspects of the solutions 
	using the collective coordinate zero-mode quantization method. 
	
\end{abstract}
	
	
	\maketitle

	\section{Introduction}
	\label{Sec.Intro}
	It is of great importance to consider $SU(3)$ generalizations of the $O(3)$  nonlinear $\sigma$-model, because they 
	may possibly play  crucial roles in  relevant limits of fundamental theories  --- for example, the 
	low-energy limit of the $SU(3)$ pure Yang-Mills theory  and the continuum limit of the $SU(3)$ Heisenberg models. 
	The main achievement of the present paper is that we have successfully constructed novel soliton solutions, 
	 called ``Hopfions," on the flag manifold $F_2=SU(3)/U(1)^2$.
	Hopfions are topological solitons with knotted structures characterized by a Hopf invariant.
	 Such knotted structures appear in various branches of physics: 
	quantum chromodynamics (QCD) \cite{Faddeev:1996zj,Faddeev:1998eq,Faddeev:1998yz}, 
	Bose-Einstein condensates \cite{Babaev:2001zy,Kawaguchi:2008xi}, 
	superconductors \cite{Babaev:2001jt}, 
	liquid crystals, \cite{Ackerman:2017lse} and so on.
	
	A typical theory that includes Hopfions is the Skyrme-Faddeev-Niemi (SFN) model 
	\cite{Faddeev:1975tz,Faddeev:1998eq},
	which is an $O(3)$ nonlinear $\sigma$-model in $(3+1)$-dimensional Minkowski space-time. 
	The scalar field theory for this model is defined by the Lagrangian density  
	\begin{equation}
	\Lag = M^2 \pd_\mu \vecn\cdot \pd^\mu \vecn-\frac{1}{2e^2}\(\pd_\mu \vecn \times \pd_\nu \vecn \)^2
	\label{Lag SF}
	\end{equation}
	where $M$ has the dimension of mass, $e$ is a dimensionless coupling constant, and $\vecn$ 
	is a three-component vector of unit length; i.e., $\vecn\cdot\vecn=1$.
	The second term on the right-hand side in \eqref{Lag SF}, 
	the Skyrme term, was introduced by Faddeev \cite{Faddeev:1975tz}  in order for 
	the theory  to satisfy Derrick's criteria for  the existence of stable soliton solutions. 
	Solutions of toroidal shape, which have the 
	lower Hopf numbers $H=1$ or $2$, were first found under an axially 
	symmetric ansatz by Gladikowski and Hellmund \cite{Gladikowski:1996mb}, and  by Faddeev and Niemi \cite{Faddeev:1996zj}.
 Hopfions with higher charge  --- including  twisted tori, linked loops, 
	and knots --- were subsequently constructed by means of full 3D energy minimization \cite{Battye:1998pe,Battye:1998zn,Hietarinta:1998kt,Hietarinta:2000ci,Sutcliffe:2003vv}.
	
	By means of the Cho-Faddeev-Niemi-Shabanov decomposition, Faddeev and Niemi 
	 showed in detail that
	the  SFN model \eqref{Lag SF} can be derived as an effective theory  that
	describes the confinement phase of the $SU(2)$ pure Yang-Mills theory \cite{Faddeev:1998eq}. 
	From  this point of view,  Hopfions are considered as  natural candidates for  glueballs  that
	can be interpreted as closed fluxtubes.
	This model is sometimes referred to as the $CP^1$ SFN model, because it is based on a formula 
	for the Lagrangian that can be described in terms of a complex scalar field via the stereographic projection $S^2\to CP^1$; i.e.,
	\begin{equation}
	\vecn=\frac{1}{\Delta}\(u+u^*,-i\(u-u^*\),|u|^2-1\)
	\label{stereographic projection}
	\end{equation}
	where $u$ is a complex scalar field and $\Delta=1+|u|^2$.	
	For a finite-energy configuration,
	the field $\vecn$ has to approach a constant vector at  spatial infinity.
	This makes the points at  infinity  identical, and the space $\Real^3$ is compactified to $S^3$.
	The field $\vecn$ defines a mapping $S^3\to S^2$, and the field configurations are characterized by an integer, 
	called a Hopf invariant, that corresponds to an  element of $\pi_3(S^2)=\integer$.
	Since  this invariant is nonlocal, an integral form of the invariant cannot be written in terms of 
	$\vecn$ or $u$;    
	in order to define it we need introduce the complex vector 
	$\vecZ=\(\Z_0,\Z_1\)^T$, with $|\vecZ|^2=1$, which satisfies $u\equiv \Z_1/\Z_0$.
	Then, the Hopf invariant can be defined as
	\begin{equation}
	H_{CP^1}=\frac{1}{4\pi^2}\int \A\wedge\dd\A ,\quad~ \A=i\vecZ^\dagger\dd\vecZ.
	\label{H SF}
	\end{equation}
	
	In this paper, we construct Hopfions in a generalization of the SFN model for the case of $SU(3)$, the gauge group of QCD.
	For $SU(N+1)$,  where $N\geq2$, there are several possibilities for the field decomposition associated 
	with dynamical symmetry-breaking patterns.
	Most studies of field decomposition are based on the following two options: 
	the maximal case $SU(N+1)\to U(1)^N$ \cite{Faddeev:1998yz} and the minimal case $SU(3)\to U(2)$ \cite{Kondo:2008xa,Evslin:2011ti}.
	Depending upon the options chosen, 
	 SFN-type models have been proposed on both the relevant target spaces, 
	$F_N=SU(N+1)/U(1)^N$ and $CP^N=SU(N+1)/U(N)$, 
	in \cite{Faddeev:1998yz} and \cite{Ferreira:2010jb}, respectively. 
	Note that $CP^1=F_1=SU(2)/U(1)$ is equivalent to $S^2$, the target space of the standard SFN model. 
	However, note also that the $CP^N~(N\geq 2)$ model cannot possess knot solitons 
	as a static stable solution in three-dimensional space, because  the  corresponding
	homotopy group  is trivial; i.e., $\pi_3(CP^N)=0$ for $N\geq2$. 
	In $2(N+1)$-dimensional space-time,  the existence of Hopfions associated 
	with $\pi_{2N+1}\(CP^{N}\)=\integer$ is discussed in \cite{Radu:2013hca}.
	In addition, if $N$ is odd, then 3D, time-dependent, non-topological solitons --- called
	$Q$-balls and $Q$-shells --- are obtained in a $CP^N$ model with a V-shaped potential \cite{Klimas:2017eft}.
	 
	Contrary to the case of $CP^2$, the third homotopy group of the flag manifold is nontrivial; i.e., $\pi_3(F_2)=\integer$. 
	Thus, we  expect Hopfions to exist in the $F_2$ SFN model, 
	which is composed of the $F_2$ \nls with quadratic terms in the derivatives. 
	The main purpose of the present paper is to confirm the existence of the $F_2$ Hopfions and 
	understand their detailed structures.	
	It has recently been found that the 2-dimensional $F_2$ \nls possesses vortex-like solutions (2D instantons), 
	both of the embedding type \cite{Ueda:2016} and of the genuine (non-embedding) type \cite{Bykov:2015pka,Amari:2017qnb}.
	The Hopfions considered in this paper are probably the vortices with knot structures. 	
	 In \cite{Bykov:2015pka}, Bykov introduced the so-called Kalb-Ramond field  
	and found that the model is integrable 
	for specific coefficients of the field \cite{Bykov:2014efa,Bykov:2016rdv}.
	Though the Kalb-Ramond field appears naturally for some continuum limits 
	of the $SU(3)$ antiferromagnetic spin chain, 
	for the moment we do not consider this field. 
	The reason is that, if one derives $F_2$ nonlinear $\sigma$-models from other fundamental theories, 
	it is  unclear whether the field can  appear naturally.	
	The genuine solutions are constructed by using the $CP^2$ Din-Zakrzewski tower generated by the \Backlund transformation \cite{Din:1980jg} 
	which implies that the solution is a composite of solitons and anti-solitons.
	
	The $F_2$ \nls has been derived from the $SU(3)$ antiferromagnetic Heisenberg model as an effective model
	on a square lattice \cite{Bernatska:2009}, 
	a triangular lattice \cite{Shannon:2013}, and a 1D chain \cite{Bykov:2011ai}. It  
	has also been  proposed to 
	describe some phenomena of a quantum spin-nematic system \cite{Shannon:2013}  and of 
	a color superconductor in high-density quark matter \cite{Kobayashi:2013axa}.  
	The present work can thus be applied to several areas of
	condensed matter physics as well as to QCD.

	 This paper is organized as follows. 
	In Sec.\ref{Sec.model} we introduce the model and some important quantities, together with a particularly 
	nice parametrization that makes the computations transparent.
	In  Sec.\ref{Sec.equation}, we derive the formal Euler-Lagrange equation, which 
	we solve  with some ansatz.
	 We give a brief analysis of some quantum aspects of the solutions  in  Sec.\ref{Sec.quantization}, and we
	conclude with Sec.\ref{Sec.conclusion}.

	\section{The model}
	\label{Sec.model}
	
	\subsection{The static energy, topological charge and torsion}
	\label{SubSec.energy}
	
	The fundamental degrees of freedom of $F_2$ nonlinear $\sigma$-models are given by 
	$su(3)$-valued fields, called ``color-direction fields" in the context of QCD. 
	They are defined by
	\begin{equation}
		\n_a=Uh_aU^\dagger, \qquad a=1,2
	\end{equation}
	where $U$ is an element of $SU(3)$, and the matrices $h_a$ are the Cartan generators in $su(3)$.
	The $F_2$ SFN model is defined by the following Lagrangian density in (3+1)-dimensional Minkowski space-time \cite{Faddeev:1998yz}:
	\begin{equation}
		\Lag=
		\sum_{a=1}^{2}\left\{
		M^2\<\pd_\mu\n_a, \pd^\mu\n_a\>-\frac{1}{e^2}F^a_{\mu\nu}{F^a}^{\mu\nu}
		\right\}
		\label{Lag}
	\end{equation}
	where the angle brackets denote the inner product on $su(3)$;
	i.e. $\<A,B\>=\Tr\(A^\dagger B\)$ for $A,B \in su(3)$.
	The second-rank tensors are defined as
	\begin{equation}
		F^a_{\mu\nu}=-\frac{i}{2}\sum_{b=1}^{2}\<\n_a,\[\pd_\mu\n_b,\pd_\nu\n_b\]\>
		\label{Kirillov-Kostant}
	\end{equation}
	and the 2-forms $F^a=\frac{1}{2}F^a_{\mu\nu} dx^\mu \wedge dx^\nu$ are called the Kirillov-Kostant (KK) symplectic forms.
	The Lagrangian \eqref{Lag} is invariant under the left global $SU(3)$ transformation  $U\rightarrow gU,~ g\in SU(3)$, 
	and the local $U(1)^2$ transformation $U\rightarrow Uk,~ k\in U(1)^2$. 
	From these symmetries, one can understand that the target space of this model 
	is the coset space $SU(3)/U(1)^2$, which is equivalent to the flag manifold $F_2$.

	The static energy functional associated with \eqref{Lag} is given by
	\begin{equation}
		E=\int d^3x~\sum_{a=1}^{2} \left\{
		\<\pd_i\n_a, \pd_i\n_a\>+F^a_{ij}F^a_{ij}
		\right\}\,.
		\label{Ene n}
	\end{equation}
	where for simplicity we use the length unit $(Me)^{-1}$ and the energy unit $4M/e$.
	Since the energy consists of both quadratic and quartic terms,
	three dimensional particle-like configurations evidently evade Derrick's no-go theorem.
	
	We reformulate the energy functional \eqref{Ene n} into a more tractable form that is given solely
	in terms of  the off-diagonal components of the Maurer-Cartan form $U^\dagger\pd_\mu U$.
	We decompose the Maurer-Cartan form in terms of the $SU(3)$ Cartan-Weyl basis as
	\begin{equation}
	U^\dagger\pd_\mu U=iA^a_\mu h_a+iJ^{p}_\mu e_p
	\label{U decomposition}
	\end{equation}
	where we use a basis of the form
	\begin{align*}
	&h_1=\frac{1}{\sqrt{2}}\lambda_3,~~h_2=\frac{1}{\sqrt{2}}\lambda_8,\\
	&e_{\pm 1}=\frac{1}{2}\(\lambda_1\pm i\lambda_2\),~
	e_{\pm 2}=\frac{1}{2}\(\lambda_4\mp i\lambda_5\),~
	e_{\pm 3}=\frac{1}{2}\(\lambda_6\pm i\lambda_7\).
	\end{align*}
	Since the basis set is orthonormal, the currents can be written as
	\begin{align}
		A^a_\mu=-i\<h_a,U^\dagger\pd_\mu U\> ,\quad
		J^p_\mu=-i\<e_p,U^\dagger\pd_\mu U\> .
		\label{A J def}
	\end{align}
	Note that $A^a_\mu$ are real and $J^{-p}_\mu=\(J_\mu^p\)^*$. 
	Under the gauge transformation $U\to Uk$, with $k=\exp\(i\theta^ah_a\)$,
	$A_\mu^a$ transforms as a gauge field and $J^p_\mu$ as a charged particle; i.e.,
	\begin{equation}
		A_\mu^a\to A_\mu^a+\pd_\mu \theta^a,\qquad
		J_\mu^p\to  J_\mu^pe^{-i\theta^a\alpha_a^p}
		\label{AJTransf}
	\end{equation}
		where $\alpha^p_a$ is  the $a$-th component of the root vector corresponding to $e_p$. Now the root vectors are given by
	\begin{equation}
	\bm{\alpha}^{1}=\(\begin{array}{c}
	\sqrt{2}\\ 
	0
	\end{array} \),
	~
	\bm{\alpha}^{2}=\frac{-1}{\sqrt{2}}\(\begin{array}{c}
	1\\ 
	\sqrt{3}
	\end{array} \),
	~
	\bm{\alpha}^{3}=\frac{1}{\sqrt{2}}\(\begin{array}{c}
	-1\\ 
	\sqrt{3}
	\end{array} \)
	\end{equation}
	with $\bm{\alpha}^{-p}=-\bm{\alpha}^p$ for $p=1,2,3$. 
	  	
	For the nonlinear $\sigma$-model, the quadratic term in \eqref{Lag},  
	can be written using only the off-diagonal components of $J^p_\mu$. 
	In addition, one can write the KK forms  as $F^a=\dd A^a=-i\sum_p \alpha^p_aJ^p \wedge J^{-p}$ where $A^a=A^a_\mu ~\dd x^\mu$, and $J^p=J^p_\mu ~\dd x^\mu$.
	Thus the static energy can be written  as 
	\begin{align}
	E=\int d^3x\sum_{q=1}^{3}\[J^q_iJ^{-q}_i
	-\frac{1}{4}\(J^q_{[i}J^{-q}_{j]}-J^{q+1}_{[i}J^{-(q+1)}_{j]}\)^2\]
	\label{Ene J}
	\end{align}
	where $q$ is a mod 3 number; i.e. $q\equiv q+3$ (mod 3).
	Note that $J^p_{[i}J^{-p}_{j]}\equiv J^p_{i}J^{-p}_{j}-J^p_{j}J^{-p}_{i}$ is purely imaginary, 
	and  therefore the energy functional is positive definite.
	It is worth  noting that, similar to the $CP^1$ case \cite{vanBaal:2001jm}, 
	the energy functional \eqref{Ene J} can be interpreted as a gauge-fixing functional for a nonlinear 
	maximal Abelian gauge, without making the Abelian subgroup components fixed.

	To ensure the finiteness of the energy functional the fields $\n_a$  must
	approach  constant matrices at  spatial infinity, so that
	the space $\Real^3$ is topologically compactified to $S^3$, and the fields $\n_a$ define the 
	map $S^3\to F_2=SU(3)/U(1)^2$.
	Consequently, the finite energy configurations can be characterized by elements of the homotopy group $\pi_3\(SU(3)/U(1)^2\)=\integer$.
	The corresponding topological charge, the Hopf invariant, is given by
	\begin{align}
		H_{F_2}=\frac{1}{8\pi^2}
		\int d^3x 
		\left\{
		\varepsilon^{ijk} \sum_{a=1}^{2}A^a_iF^a_{jk}
		\right\}-\Gamma
		\label{Hopf charge}
	\end{align}
	where
	\begin{align}
		\Gamma=\frac{-i}{8\pi^2}\int d^3x~\varepsilon^{ijk}\left\{
		J^1_iJ^2_jJ^3_k-J^{-1}_iJ^{-2}_jJ^{-3}_k
		\right\}\,.
	\end{align}
	The Hopf invariant \eqref{Hopf charge} is nonlocal, since $A^a_\mu$ cannot be written in terms of the fields $\n_a$, 
	and therefore \eqref{Hopf charge} does not possess  local $U(1)^2$ symmetry.
	Note that both the Abelian Chern-Simons (CS) terms and $\Gamma$ are not topological.
	The Hopf invariant can be constructed by means of  Novikov's procedure \cite{Kisielowski:2013ina} 
	via the isomorphism between $\pi_3\(SU(3)/U(1)^2\)$ and $\pi_3\(SU(3)\)$, 
	which indicates  $H_{F_2}=Q[U]$, where $Q[U]=\frac{1}{24\pi^2}\int\Tr\(U^\dagger\dd U\)^3$ is the winding number of the map $U : S^3\to SU(3)$.
	
	The winding number is equivalent to the  CS term for the $SU(3)$ flat connection $U^\dagger\dd U$. 
	Therefore, similar to the $CP^1$ Hopf invariant \eqref{H SF} discussed in \cite{Langmann:1999nn,vanBaal:2001jm}, the $F_2$ Hopf invariant \eqref{Hopf charge} can also be given by the non-Abelian  CS term with the $SU(3)$ flat connection.

	 The \Kahler form can be defined as
	\begin{equation}
		\lambda=\frac{i}{2\pi}\sum_{p=1}^{3}B_pJ^p\wedge J^{-p}
	\end{equation}
	where the coefficients $B_p$ are real constants \cite{Bykov:2014efa}.
	For non-symmetric manifolds, like the flag space $F_2$, the \Kahler form $\lambda$  in general is not closed, i.e., 
	$\dd \lambda\neq 0$.
	The so-called skew torsion $T=\dd \lambda$ is given by the form
	\begin{align}
		T=\frac{1}{2\pi}\sum_p B_p\(J^1\wedge J^2\wedge J^3+J^{-1}\wedge J^{-2}\wedge J^{-3}   \)\,.
	\end{align}
	Under the local $U(1)^2$ transformation, both the \Kahler form  and the torsion are invariant.
	Note that in the 2-dimensional $F_2$ nonlinear $\sigma$-model, the solutions of the Euler-Lagrange equation 
	make the \Kahler form  closed, and  the torsion then disappears identically.
	By analogy, in this paper we consider a class of configurations that satisfies the torsion-free condition $T=0$.

	\subsection{Parametrization}

	In order to make the analysis transparent, let us parametrize the $SU(3)$ matrix $U$ in terms of complex scalar fields which are
	equivalent to the local coordinates of the target space $F_2$.	
	The coordinates can be introduced naturally via the inverse of a generalization of stereographic projection; i.e., by 
	the mapping $SL(3,\Complex)/B_+\to SU(3)/U(1)^2$, where $B_+$ is the Borel subgroup of upper triangular matrices (see, e.g., \cite{Picken:1988fw}).
	However, we need two additional degrees of freedom in order to describe the Hopf invariant, because it is nonlocal and requires $8=\dim SU(3)$ degrees of freedom rather than $6=\dim F_2$.
	Therefore, we begin the parametrization not with a $3\times 3$ lower triangular matrix in which all the diagonal components are unity
	which is an element of $SL(3,\Complex)/B_+$ but instead with a lower triangular matrix in $SL(3,\Complex )$ of the form
	\begin{equation}
		X=\left(\begin{array}{ccc}
			\chi_1 & 0 & 0 \\ 
			\chi_2 & \chi_4 & 0 \\ 
			\chi_3 & \chi_5 & \(\chi_1\chi_4\)^{-1}
		\end{array} \right)\, 
		 \in SL(3,\Complex)
		\label{low tri}
	\end{equation}
	where the $\chi_i$ are complex functions, with $\chi_1$ and $\chi_4$ being finite.
	Note that the matrix \eqref{low tri} has ten degrees of freedom.
    
    The parametrization can then be obtained by using the Gram-Schmidt orthogonalization process.  
	We write $X$ in terms of column vectors as $X=(\vec{c}_1,\vec{c}_2,\vec{c}_3)$, and 	
	we introduce the mutually orthogonal vectors $\vec{v}_j$:
	\begin{equation}
	\begin{split}
	& \vec{v}_1=\vec{c}_1~,
	\\
	& \vec{v}_2=\vec{c}_2-\frac{(\vec{c}_2,\vec{v}_1)}{(\vec{v}_1,\vec{v}_1)}\vec{v}_1~, 
	\\
	&
	\vec{v}_3=\vec{c}_3-\frac{(\vec{c}_3,\vec{v}_2)}{(\vec{v}_2,\vec{v}_2)}\vec{v}_2-\frac{(\vec{c}_3,\vec{v}_1)}{(\vec{v}_1,\vec{v}_1)}\vec{v}_1~.
	\end{split}
	\label{vector e}
	\end{equation}
	Normalization of the vectors $\vec{v}_j$ is achieved under the two conditions 
	\begin{equation}
		\begin{split}
			&|\chi_1|^2+|\chi_2|^2+|\chi_3|^2=1 , \\
			&|\chi_1|^2\(|\chi_4|^2+|\chi_5|^2\)+|\chi_3\chi_4-\chi_2\chi_5|^2=1\,.
		\end{split}	
		\label{normalizing condition}
	\end{equation}
	Then we write  
		\begin{equation}
			U=\(\vec{v}_1,\vec{v}_2,\vec{v}_3\) .
			\label{U-v}
		\end{equation}
	This is a unitary matrix with eight degrees of freedom, because the vectors $\vec{v}_j$ form a complete basis set, and they 
	are described by the five complex scalars $\chi_i$ with the two constraints \eqref{normalizing condition}.	
	Finally, we parametrize the $\vec{v}_i$ in terms of three complex scalar fields, which correspond to the local coordinates of the flag manifold. 	
	We introduce  them as
	$(u_1,u_2,u_3)=(\chi_2/\chi_1,~\chi_3/\chi_1,~\chi_5/\chi_4)$, where we
	also write $\arg\(\chi_\alpha\)=\vartheta_\alpha$ for $\alpha=1,4$. Then the $SU(3)$ matrix \eqref{U-v}
	 can be written as 
	$U=\(
	\ZA e^{i\vartheta_1},
	\ZB e^{i\vartheta_4},
	\ZC e^{-i(\vartheta_1+\vartheta_4)}\)$
	where
	\begin{equation}
		\begin{split}
			Z_\rmA&=
			\frac{1}{\sqrt{\Delta_1}}
			\left(\begin{array}{c}
				1\\ 
				u_1\\ 
				u_2
			\end{array} \right),
			\\
			Z_\rmB&=
			\frac{1}{\sqrt{\Delta_1\Delta_2}}
			\left(\begin{array}{c}
				-u_1^*-u_2^*u_3\\
				1-u_1u_2^*u_3+|u_2|^2\\
				-u_1^*u_2+u_3+u_3|u_1|^2 
			\end{array} \right),
			\\
			Z_\rmC&=
			\frac{1}{\sqrt{\Delta_2}}
			\left(\begin{array}{c}
				u_1^*u_3^*-u_2^*\\
				-u_3^*\\
				1 
			\end{array}
			\right)
		\end{split}
		\label{explict Z}
	\end{equation}
	 with 
	\begin{equation}
		\begin{split}
			\Delta_1&=1+|u_1|^2+|u_2|^2~,
			\\
			\Delta_2&=1+|u_3|^2+|u_1u_3-u_2|^2.
		\end{split}
		\label{formdelta}
	\end{equation}	
    The unitarity of $U$ is guaranteed by the orthonormalization condition and the completeness relation between the complex vectors:
	\begin{gather}
	Z_a^\dagger Z_b=\delta_{ab}~,   \label{orthonormal}
	\\
	Z_\rmA\otimes Z_\rmA^\dagger
	+Z_\rmB\otimes Z_\rmB^\dagger
	+Z_\rmC\otimes Z_\rmC^\dagger=\mathbf{1}_3\,. \label{completeness}
	\end{gather}
	These identities are helpful for all the computations in this study.
	Note that the triplet $\{\ZA, \ZB, \ZC\}$ plays the role of an order parameter 
	if the model is viewed as an effective theory of the $SU(3)$ antiferromagnetic Heisenberg model \cite{Shannon:2013,Ueda:2016}.
	
	As mentioned earlier, we require eight degrees of freedom to describe the Hopf invariant. However, 
	the two degrees of freedom corresponding to $\vartheta_1$ and $\vartheta_4$ are canceled out in the energy, 
	the Euler-Lagrange equation and so on, although these variables make the calculations much more complicated. 
	Therefore it is useful to introduce the matrix $W$ without the phase factors; i.e.,  
	\begin{equation}
		W=\(\ZA,\ZB,\ZC\).
	\end{equation}
	This can also be written as $W=U\exp\[i\Theta^ah_a\]$ with $\Theta^1=-\frac{\vartheta_1-\vartheta_4}{\sqrt{2}}$ and $\Theta^2=-\frac{\sqrt{3}(\vartheta_1+\vartheta_4)}{\sqrt{2}}$.	
	By virtue of the local symmetry, the Lagrangian satisfies
	$\Lag\[\n_a\]=\Lag\[\m_a\]$ where $\m_a=Wh_a W^\dagger$.
	In addition, the static energy can be written in terms of the off-diagonal components of the Murer-Cartan form $W^\dagger\pd_\mu W$.
	We again decompose it as 
	\begin{equation}
	W^\dagger \pd_\mu W=iC^a_\mu h_a+iK^{p}_\mu e_p
	\label{W decomposition}
	\end{equation} 
	and \eqref{AJTransf} then yields the relations 
	\begin{equation}
			C^a_\mu=A^a_\mu +\pd_\mu \Theta^a,~~~
		K^p_\mu=J^p_\mu e^{-i\alpha_a^p\Theta^a}.
	\end{equation}
	
	Since $J^{-p}_\mu=(J^p_\mu)^*$, we can replace $J^p_i$ in the energy \eqref{Ene J} with $K^p_i$: The static energy thus 
can also be written as 
	\begin{align}
	E=\int d^3x\sum_{q=1}^{3}\[K^q_iK^{-q}_i
	-\frac{1}{4}\(K^q_{[i}K^{-q}_{j]}-K^{q+1}_{[i}K^{-(q+1)}_{j]}\)^2\].
	\label{Ene K}
	\end{align}
	Also, the KK form, \Kahler form and skew torsion can be written as
	\begin{align}
	&F^a=-2i\sum_{p=1}^{3}\alpha^p_aK^p\wedge K^{-p}, \\
	&\lambda=\frac{i}{2\pi}\sum_{p=1}^{3}B_pK^p\wedge K^{-p}, \\
	&T=\frac{1}{2\pi}\sum_{p=1}^3 B_p\(K^1\wedge K^2\wedge K^3+K^{-1}\wedge K^{-2}\wedge K^{-3}   \)\,. 
	\label{Torsion K}
	\end{align}
	Hereinafter we use $W$ rather than $U$, except for the Hopf invariant. 
	
	\section{Equation of motion and Hopfions}
	\label{Sec.equation}
	First we derive the formal Euler-Lagrange equation, and we then  consider 
	two classes of configurations that satisfy the torsion-free condition $T=0$.
	The Euler-Lagrange equation is equivalent to the conservation law for the Noether current $\K_\mu$ 
	associated with the global $SU(3)$ transformation; i.e., $\pd_\mu\K^\mu=0$.
	The current takes the form
	\begin{equation}
		\K_\mu=\sum_{a=1}^{2}\(\[\m_a,\pd_\mu\m_a \]-i\sum_{b=1}^{2} F^a_{\mu\nu}\[\m_a,\[\m_b,\pd^\nu\m_b\]\]\)\,.
	\end{equation}
	where $\m_a=Wh_a W^\dagger$.
	If we factorize the current as $\K_\mu=W\B_\mu W^\dagger$, the equations of motion can be written as
	\begin{align}
		\pd_\mu\B^\mu+\[W^\dagger\pd_\mu W,\B^\mu\]=0\,.
		\label{ELeq B}
	\end{align}
 The current $\B_\mu$ consists of just the off-diagonal components of the Murer-Cartan form: 
	\begin{align}
		\B_\mu =i\sum_{p}\(K^{p}_\mu-i\sum_{a=1}^{2}\alpha_a^pF^a_{\mu\nu}{K^{p}}^\nu \)e_p\,.
		\label{B def}
	\end{align}
	To simplify the notation, 
	we introduce $R^p_\mu=\sum_a\alpha^p_aC^a_\mu$ and $G^p_{\mu\nu}=\sum_a\alpha^p_aF^a_{\mu\nu}$.
	Then, equation \eqref{ELeq B} can  be written explicitly as 
	\begin{equation}
		\begin{split}
		&\pd^\mu\(K^q_\mu-iG^q_{\mu\nu}{K^q}^\nu\) 
		\\
		&\quad +i{R^q}^\mu\(K^q_\mu -iG^q_{\mu\nu}{K^q}^\nu\)+{G^q}^{\mu\nu}K^{-q-1}_\mu K^{-q+1}_\nu=0
		\end{split}
		\label{ELeq K}
	\end{equation}
	for $\forall q\equiv 1,2,3$ (mod 3) and their complex conjugations.

	The normal route to confirm the existence of knot solitons consists of solving  equation \eqref{ELeq K} with some symmetric ansatz for the complex scalar fields $u_i$.
	However, since \eqref{ELeq K} is highly nonlinear and very complicated, this seems quite a hard task. 
	Here, we instead employ a different strategy.
	That is,  we first introduce configurations that satisfy the torsion-free 
	condition $T=0$ and then, as we shall see, the Euler-Lagrange equation \eqref{ELeq K}  
	simplify into a solvable form.
	
	\subsection{Trivial $CP^1$ reduction}
	The first class we consider is a trivially embedded configuration; i.e., an $F_1=CP^1$ Hopfion into $F_2$ space.   
	It can be obtained by requiring two of the three scalar fields to be trivial.
	Without loss of generality,  we here set $u_1=u_3=0$ and  write $u_2=u$.
	Then, the complex vectors $Z_a$  can be written in terms of the function $u(x)$ as
	\begin{equation}
		\ZA=\frac{1}{\sqrt{\Delta}}\(\begin{array}{c}
		1\\ 
		0\\ 
		u
		\end{array} \),~
		\ZB=\(\begin{array}{c}
		0\\ 
		1\\ 
		0
		\end{array} \),~
		\ZC=\frac{1}{\sqrt{\Delta}}\(\begin{array}{c}
		-u^*\\ 
		0\\ 
		1
		\end{array} \)
		\label{Z trivial}
	\end{equation}
	where $\Delta=1+|u|^2$.	
	The currents $K_\mu^p$ are given by   
	\begin{equation}
		K^1_\mu=K^3_\mu=0,\qquad K^2_\mu=\frac{i}{\Delta}\pd_\mu u
	\end{equation}
	and the skew torsion $T$ vanishes. It can be checked directly that the equations of motion 
	\eqref{ELeq K} for $q\equiv 1$ and $3$ are automatically satisfied and that for $q\equiv2$ reduces to	
	\begin{equation}
	\begin{split}
	&\pd^\mu\[\pd_\mu u-iG_{\mu\nu}\pd^\nu u \]\\
	&\qquad\qquad +\(iR_\mu-\pd_\mu \log\Delta\)\(\pd^\mu u-iG^{\mu\nu}\pd_\nu u\)=0
	\end{split}
	\label{ELeq tri}
	\end{equation}
	where for convenience we have introduced $R_\mu$ and $G_{\mu\nu}$, which take the forms
	\begin{align}
	& R_\mu \equiv\frac{i}{\Delta}\(u^*\pd_\mu u-u\pd_\mu u^* \)\,,
	\label{R def}\\
	&G_{\mu\nu}\equiv-\frac{2i}{\Delta^2}\(\pd_\mu u\pd_\nu u^*-\pd_\mu u^*\pd_\nu u \)\,.
	\label{G def}
	\end{align}  
	The static energy for the configuration \eqref{Z trivial} is given by
	\begin{equation}
	E_{\tri}=\int d^3x\(\frac{\pd_i u\pd_iu^*}{\Delta^2}
	-\frac{\(\pd_i u\pd_j u^*-\pd_i u^*\pd_j u\)^2}{2\Delta^4} \)\,.
	\label{Energy tri}
	\end{equation}	
	Both the equation of motion \eqref{ELeq tri} and the energy \eqref{Energy tri} are exactly the same as those of the $CP^1$ SFN model \eqref{Lag SF} with \eqref{stereographic projection}. 
	Next we determine the Hopf invariant. First we plug the configuration \eqref{Z trivial} into the $SU(3)$ matrix $U$ and define 
	$\Z_0=e^{i\vartheta_1}/\sqrt{\Delta},~ \Z_1=ue^{i\vartheta_1}/\sqrt{\Delta}$ with $\vartheta_4=0$. Then we obtain 
\begin{equation}
A^1=-\frac{1}{\sqrt{2}}\mathcal{A}, \quad
A^2=\sqrt{\frac{3}{2}}\mathcal{A},
\end{equation} 
where
\begin{equation}
\A=i\vecZ^\dagger\dd\vecZ, \quad\vecZ=\(\begin{array}{c}
\Z_0\\ 
\Z_1
\end{array} \).
\end{equation}
Therefore we find that	 
the $F_2$ Hopf invariant \eqref{Hopf charge} of the embedding configuration coincides with the $CP^1$ version \eqref{H SF}, i.e., 
\begin{equation}
H_{\tri}=\frac{1}{4\pi^2}\int \A\wedge\dd\A .
\label{Hopf Ch tri}
\end{equation}
	This coincidence is obviously due to the fact that \eqref{Z trivial} is just a trivial embedding configuration.
	Next we examine another class of configuration which  has a more nontrivial nature, 
	and see what happens to the Euler-Lagrange  equation, the energy, and the Hopf invariant. 
	 
	\subsection{Nontrivial $CP^1$ reduction}
	
	For the trivial embedding \eqref{Z trivial}, we observed that two pairs of the currents $K^p_\mu$ vanished;
	i.e., $K^{\pm 1}_\mu=K^{\pm 3}_\mu=0$.
	Here we relax these conditions and  examine the case where just one pair of the currents vanishes; i.e., $K^{\pm2}_\mu=0$,  
	while both $K^{\pm 1}_\mu$ and $K^{\pm 3}_\mu$ remain finite. 
	This automatically satisfies the torsion-free condition.
	Note that the result is independent  of the choice of the components; 
	for a different pair, one just repeats the same prescription by permuting the vectors $Z_a$. 	
	The condition $K_\mu^{\pm2}=0$ reads
	\begin{equation}
		u_3\pd_\mu u_1-\pd_\mu u_2=0, \qquad \mu=0,1,2,3.
	\end{equation}
	 This is satisfied if $u_2$ is a function of $u_1$ --- i.e., $u_2=f(u_1)$ ---
	and  $u_3$ is given by $u_3=f^\prime(u_1)$ 
	where the prime stands for the derivative with respect to $u_1$.
	This means that the only independent field is  $u_1$, 
	so that the Euler-Lagrange equation seems to be an overdetermined system.
	In order to  reduce the number of independent equations, 
	we consider the case where the Euler-Lagrange equations for $q\equiv 1$ and  $3$ are proportional to each other.
	This is the case when the ratio $\DOne/\DTwo$ is a constant.
	Note that we leave the equation for $q\equiv\pm2$ intact because $q\equiv2$ is now special due to the constraint $K^2_\mu=0$.	 
	By comparing the order of $u_1$ in $\Delta_i$'s, one finds that this condition requires
	\begin{equation}
		|u_1|^2=|f^\prime(u_1)|^2,\qquad
		|f|^2=|u_1f'-f|^2\,.
	\end{equation}
	Since we are not interested in embedding solutions here, we omit the case where $u_1$ is a constant 
	and obtain
	\begin{equation}
		f(u)=\frac{1}{2}u^2e^{i\varphi}
	\end{equation}
	where $\varphi\in[0,2\pi]$ is a constant.
	Note that due to $U(1)$ symmetries, the constant $\varphi$ can take an arbitrary value. 
	For simplicity, we choose $\varphi=\pi$ and write $u_1=\sqrt{2}u$.
	Then, the triplet vectors  become
	\begin{equation}
		\begin{split}
		&\ZA=\frac{1}{\Delta}\(1,\sqrt{2}u,-u^2\)^T,~
		\ZC=\frac{1}{\Delta}\(-{u^*}^2,\sqrt{2}u^*,1\)^T,\\
		&\qquad\qquad\quad
		\ZB=\frac{1}{\Delta}\(-\sqrt{2}u^*,1-|u|^2,-\sqrt{2}u\)^T\,.
		\end{split}
		\label{Z nontri}
	\end{equation}
	It is worth  noting that the three vectors are linked by the \Backlund transformation, i.e.,
	\begin{equation}
		\ZB=\frac{P_+\ZA}{|P_+\ZA|},\qquad \ZC=\frac{P_+\ZB}{|P_+\ZB|}
	\end{equation}
	where $P_+Z_a=\pd_uZ_a-\(Z_a^\dagger\pd_u Z_a\)Z_a$.
	Similar relations among the triplet vectors are  observed 
	for the non-embedding solutions of the two-dimensional $F_2$ nonlinear $\sigma$-model \cite{Bykov:2015pka,Amari:2017qnb}. 	
	The currents $K_\mu^p$ are given by the form
	\begin{equation}
		K^1_\mu =\frac{\sqrt{2}i}{\Delta}\pd_\mu u^*,\quad 
		K^2_\mu=0,\quad
		K^3_\mu =-\frac{\sqrt{2}i}{\Delta}\pd_\mu u^*.
		\label{k123}
	\end{equation}
	With the forms \eqref{k123}, the Euler-Lagrange equation \eqref{ELeq K} for $q\equiv2$ is automatically satisfied. 
	In addition, we obtain $R^1_\mu=R^3_\mu=-R_\mu$ and $G^1_{\mu\nu}=G^3_{\mu\nu}=-G_{\mu\nu}$,  and one then 
	 observes easily that  equations \eqref{ELeq K} for both $q\equiv 1$ and $3$  reduce 
	to the complex  conjugates of \eqref{ELeq tri}. 
	To see this, one can use the fact that $R_\mu$ and $G_{\mu\nu}$ are real. 
	This yields a somewhat surprising observation: 
	These results clearly mean that it is not only in the trivial embedding case but also in the non-embedding case that 
	all the known Hopfion solutions $u$ in the $CP^1$ SFN model solve the Euler-Lagrange equation.

	Though the Euler-Lagrange equations in both  classes are solved by the same function, they are clearly inequivalent.   
	The configuration \eqref{Z nontri} possesses a static energy 
	that is exactly four times greater than \eqref{Z trivial}; i.e.,
	\begin{equation}
		E_\nontri[u]=4E_\tri[u].
		\label{Enef2}
	\end{equation}
	To evaluate the relevant Hopf number, 
	we  write $u=\Z_1/\Z_0,~e^{i\vartheta_1}=\Z_0^2/|\Z_0|^2$, and 
	$ \vartheta_4=0$.
	This yields
	$A^1=-\sqrt{2}\A,~ A^2=\sqrt{6}\A$, and therefore
	\begin{equation}
		H_\nontri=\frac{1}{\pi^2}\int\A\wedge\dd\A=4H_\tri.
	\label{Hopff2}
	\end{equation}
	It is worth noting that the $F_2$ Hopf invariant is equivalent to a non-Abelian CS term 
	with an $SU(3)$ flat connection, but that of the solutions  is  given by 
	the sum of just the Abelian CS terms, 
	because the configuration \eqref{Z trivial} satisfies $\Gamma=0$.  

	According to \eqref{Enef2} and \eqref{Hopff2}, the $F_2$ nontrivial Hopfion with $H_\nontri=4n$ (where $n$ is an integer)  can be 
	viewed as  a molecular state of four embedding solutions with the Hopf 
	number $H=n$  which are 
	sitting on top of each other with no binding energy.
	Such a situation has been observed in an $SU(N)$ Skyrme model~\cite{Ioannidou:1999mk}. 
	Since our solutions of equation \eqref{ELeq tri} are not of the BPS type, 
	it is probably impossible to remove one of them from the others without changing the energy. 
	However, this should be confirmed by studying the moduli parameters of the solutions. This will be reported in subsequent papers.

	\section{Iso-spinning Hopfions}
	\label{Sec.quantization}
	
	We have seen that the Euler-Lagrange equations are solved by the same function $u$
	both for  the trivial embedding and for the non-embedding ansatz.  
	Their energies and Hopf invariants are respectively proportional, as shown in  \eqref{Enef2} and \eqref{Hopff2}.
	In the previous sections, we saw that they are inequivalent, although they look similar. Here we shall show that their quantum natures are quite different. 
	In this section, we give a brief analysis to demonstrate the notable differences 
	in some quantum aspects based on the collective coordinate quantization
	of the zero modes. 
	We consider an adiabatic iso-rotation associated  with the $SU(3)$ global symmetry, i.e.,
	with the time-dependent transformation $\m_a(\vec{x})\to \m_a(t,\vec{x})=\beta(t)\m_a(\vec{x})\beta^\dagger(t)$,
	where $\beta (t) \in SU(3)$. 
	Then the Lagrangian can be written as
	\begin{equation}
	\begin{split}
	L=&-E_\cl\\
	&+r_0^2\int d^3x
	\[
	\Tr\(\[\bdb,\m_a\]\[\bdb,\m_a\]\)+2F^a_{0i}F^a_{0i}
	\]
	\end{split}
	\label{qLag}
	\end{equation}	
	where $E_\cl$ is the static energy of the Hopfion, the dot denotes the time derivative --- i.e., 
	$\dot{\beta}=d\beta/dt$ --- and 
	\begin{equation}
	F^a_{0j}=-\frac{i}{2}\Tr\(\m_a\[\[\bdb,\m_b\],\pd_j\m_b\]\).
	\end{equation}
	The energy collection depends on the  length scale  $r_0=\(Me\)^{-1}$.
	
	In order for the integral in \eqref{qLag} to be finite, $\bdb$ and $\m_a$ must commute with each other at spatial infinity.
	Since the fields $\m_a(\vec{x})$ approach constant elements of $u(1)\times u(1)$ as $x$ goes to infinity, $\bdb$ must also be in $u(1)\times u(1)$ and therefore can be written as
	\begin{equation}
		\bdb=\sqrt{2}i\(\frac{\omega_1}{2} h_1+\frac{\omega_2}{\sqrt{3}}h_2\)
		\label{angular velocity}
	\end{equation}
	where $\omega_a$ denotes the angular velocity in  iso-space.
	We chose the coefficients in \eqref{angular velocity} to be consistent with the definition of the $SU(3)$ Euler angle \cite{Nelson:1967eu}. 
	
	The quantum Lagrangian \eqref{qLag} can be written as a quadratic form of the angular velocities, 
	\begin{equation}
	L=-E_\cl+\frac{1}{2}\veco^T \mathcal{I}\veco
	\end{equation}
	where $\veco^T=\(\omega_1,\omega_2\)$ and 
	\begin{equation}
	\mathcal{I}=\(\begin{array}{cc}
	I_{11}&I_{12}  \\ 
	I_{12}&I_{22} 
	\end{array} \).
	\end{equation}
	The moments of inertia are explicitly obtained as follows:
	\\
	\textit{$\bullet$ For the non-trivial reduction case,} 
	\begin{align}
	\begin{split}
	\begin{split}
	&I_{11}=2r_0^2\int d^3x~\frac{1}{\Delta^4}\[\(10-7|u|^2+10|u|^4\)|u|^2 \right. \\
	&\left.\qquad\qquad\qquad\quad+4\(7-13|u|^2+7|u|^4\)\(\pd_i\log\Delta\)^2\]
	\end{split}
	\\
	\begin{split}
	&I_{12}=-4r_0^2\int d^3x~\frac{1}{\Delta^4}\(3-\Delta\)\(3-2\Delta\)\\
	&\qquad\qquad\qquad\qquad\times\[|u|^2+4\(\pd_i\log\Delta\)^2\]
	\end{split}
	\\
	\begin{split}
	&I_{22}=8r_0^2\int d^3x~\frac{1}{\Delta^4}\[\(2+|u|^2+2|u|^4\)|u|^2 
	\right. \\
	&\left.\qquad\qquad\qquad\qquad
	+4\(1-|u|^2+|u|^4\)\(\pd_i\log\Delta\)^2\].
	\end{split}
	\end{split}
	\label{moi nontri}
	\end{align}
	\textit{$\bullet$ For the trivial reduction case,} 
	\begin{equation}
	I_{11}=\frac{I_{12}}{2}=\frac{I_{22}}{4}=2r_0^2\int d^3x~\frac{|u|^2+\(\pd_i\log\Delta\)^2}{\Delta^2}\,.
	\label{moi tri}
	\end{equation}
	Using a Legendre transformation of the Lagrangian \eqref{qLag}, 
	we obtain the Hamiltonian  $\mathcal{H}=\omega_iP_i-L$ with the canonical momentum defined by 
	\begin{equation}
		P_i\equiv \frac{\pd L}{\pd \omega_i}=I_{ij}\omega_j, \qquad i,j=1,2.
		\label{def P}
	\end{equation}
	In the nontrivial reduction case, the Hamiltonian is  obtained straightforwardly
	as
	\begin{equation}
		\mathcal{H}=E_\cl+\frac{1}{2}\frac{1}{ \Det \mathcal{I}}\left\{
		I_{22}P_1^2-2I_{12}P_1P_2+I_{11}P_2^2
		\right\}
		\label{H nontri}
	\end{equation}	
	where we have used the commutation relation $\[P_1,P_2\]=0$, because the operators are associated the Abelian subgroup of $SU(3)$.
	Since the operators are already diagonalized, the Hopfions can be assigned two quantum numbers associated with the two zero-modes when the Hamiltonian operates a relevant wave function.
	On the other hand, in the embedding case, we are  allowed to define only one operator because 
	the $SU(3)$ matrix $W$ satisfies the commutation relation $	\[W,h_1-\frac{1}{\sqrt{3}}h_2\]=0$
	and therefore
	\begin{equation}
		\[\dot{\beta}\beta,\m_a\]=\frac{i\(\omega_1+2\omega_2\)}{\sqrt{2}}\[h_1,\m_a\].
	\end{equation}
	This implies that the embedding Hopfions can rotate around only one axis in isospace. 
	Actually, we can obtain from \eqref{def P} only one operator, which has the form
	\begin{equation}
		P_1=\frac{P_2}{2}=I\(\omega_1+2\omega_2\)\equiv P
	\end{equation} 
	where we have written $I_{11}=I$.
	Therefore the Hamiltonian  becomes
	\begin{equation}
		\mathcal{H}=E_\cl+\frac{P^2}{2I}\,.
		\label{H embedding}
	\end{equation}	
	Consequently,  Hopfions of the embedding type inherit the quantum properties 
	of the $CP^1$ Hopfions; they can possess at most one quantum number after \eqref{H embedding} 
	acts on a proper wave function.
	The quantum properties of the two types of  Hopfion solutions seem quite different, 
	at least qualitatively, which is  a reflection of their different symmetries.   
	
	\section{Conclusion}
	\label{Sec.conclusion}
	
	We have studied Hopfions in the SFN model on the target space $F_2=SU(3)/U(1)^2$ which is an $SU(3)$ generalization of 
	the standard SFN model, for which the  target space is $CP^1=SU(2)/U(1)$.
	By analogy  with the 2-dimensional $F_2$ nonlinear $\sigma$-model, we introduced two classes of configurations
	that satisfy the torsion-free condition; i.e., a trivial 
	embedding of the $CP^1$ Hopfions and
	the $SU(3)$ genuine one, which can be constructed through the \Backlund transformation.
	For both cases, the Euler-Lagrange equation reduces to that of the $CP^1$ SFN model.
	In addition, though the Hopf invariant is equivalent to the  
	CS term for the $SU(3)$ flat connection, we showed that the invariant 
	of the solutions is also given by the  CS terms for the 
	Abelian components of the flat connection. 
	
	The most important open problem is probably the stability of the genuine solutions. 
	Their energy and Hopf invariant are exactly  four times greater than those 
	of the embeddings, comparing the two configurations given by the same scalar function. 
	On the other hand, since the embedding Hopfions are essentially equivalent to the $CP^1$ Hopfions, their energy 
	with $H=4n$ is less than four times the energy of one with $H=n$; i.e., $E_{H=4n}<4\times E_{H=n}$ \cite{Sutcliffe:2003vv}.  
	Therefore, the genuine solutions with $H=4n$ are likely to decay into the 
	embedding solution with $H=4n$, rather than into four solutions with $H=n$.
	It is well worth confirming whether or not the genuine solutions are stable, 
	and if not, what they decay into.
	Note that even if they are unstable,  these solutions probably play an important role in some branch of physics, 
	like the known saddle-point solutions: the electro-weak sphaleron, and the meron in the pure Yang-Mills theory.
	However, the stability may restrict the potential for applications.  
	It is also  important to understand the mathematical implications of 
	the torsion-free condition in detail and to confirm whether there exist Hopfions outside this condition. 
	
	We also examined  some quantum aspects of the Hopfions based on 
	collective coordinate quantization and 
	found that they are quite different for different  
	\ansatze as a result of the difference in their symmetries.  
	  
	 To determine the physical spectra of  glueballs, 
	we need to perform a more complete analysis 
	of the collective coordinate quantization, including  rotational modes, and also  to 
	discuss their statistical properties. 
	The analysis of this subject is now in progress and the results will be reported in a subsequent paper.

	\vskip 0.5cm\noindent
	\textbf{Acknowledgment}

	The authors are grateful to Prof. L. A. Ferreira for many helpful discussions and the hospitality at his institute where a part of this work was done.
	We would like to thank D. Bykov, Y. Akagi, K. Toda and A. Nakamula for useful discussions. 
	This study was supported by JSPS KAKENHI Grant Number JP18J12275.



%

\end{document}